\documentclass[12pt]{article}

\usepackage{amsmath,amssymb,amsfonts,latexsym}
\usepackage{bm}
\usepackage{amsmath}
\newcommand{\be}{\begin{enumerate}}
\newcommand{\ee}{\end{enumerate}}
\newcommand{\ba}{\begin{array}}
\newcommand{\ea}{\end{array}}
\newcommand{\beq}{\begin{equation}}
\newcommand{\eeq}{\end{equation}}
\newcommand{\bqa}{\begin{eqnarray}}
\newcommand{\eqa}{\end{eqnarray}}
\newcommand{\bqas}{\begin{eqnarray*}}
\newcommand{\eqas}{\end{eqnarray*}}

\begin{document}

\newtheorem{defi}{Definition}[section]
\newtheorem{lem}[defi]{Lemma}
\newtheorem{prop}[defi]{Proposition}
\newtheorem{theo}[defi]{Theorem}
\newtheorem{rem}[defi]{Remark}
\newtheorem{cor}[defi]{Corollary}

\newcommand{\qed}{\hfill $\Box$\vspace{.5cm}\medskip}


\title{Study of Non-Holonomic Deformations of Non-local integrable systems belonging to the Nonlinear Schr\"odinger family}

\author {Indranil Mukherjee$^1$\footnote{E-mail: {\tt indranil.m11@gmail.com}} and Partha Guha$^2$\footnote{E-mail: {\tt partha@bose.res.in}} \\$^1$ School of Natural and Applied Sciences,\\ Maulana Abul Kalam Azad University of Technology,\\ West Bengal, India\\$^2$S.N. Bose National Centre for Basic Sciences,\\ JD Block, Sector III, Salt Lake, Kolkata - 700098,  India \\
}

\date{}
\maketitle

\hspace{1.20 in}

\vspace{.15 in}

\abstract{The non-holonomic deformations of non-local integrable systems belonging to the Nonlinear Schr\"odinger family are studied using the Bi-Hamiltonian formalism as well as the Lax pair method. The non-local equations are first obtained by symmetry reductions of the variables in the corresponding local systems. The bi-Hamiltonian structures of these equations are explicitly derived. The bi-Hamiltonian structures are used to obtain the non-holonomic deformation following the Kupershmidt ansatz. Further, the same deformation is studied using the Lax pair approach and several properties of the deformation discussed. The process is carried out for coupled non-local Nonlinear Schr\"odinger and Derivative Nonlinear Schr\"odinger (Kaup Newell) equations. In case of the former, an exact equivalence between the deformations obtained through the bi-Hamiltonian and Lax pair formalisms is indicated.

\bigskip



\bigskip

{\bf Keywords and Keyphrases :} Non-local integrable systems, Nonlinear Schr\"odinger equation, Kaup-Newell equation, bihamiltonian system,
Lax method, nonholonomic deformation.

\section{Introduction}
The importance of completely integrable systems in different areas such as water waves, plasma physics, field theory, nonlinear optics, lattice dynamics etc. can hardly be overemphasized \cite{Das}. The standard technique to study integrable models is by using the Lax pair, using the zero curvature or flatness condition \cite{Lax}. Systems are considered to be integrable when they contain infinitely many conserved quantities which lead to the stability of the soliton solutions. These constants of motion determine the system dynamics, thereby allowing solution by the method of Inverse Scattering Transform (IST) in appropriate variables \cite{AC, AKNS}. The integrable systems are found to possess a local bi-Hamiltonian structure \cite{FT, Magri}. It is well known that starting from a suitably chosen spectral problem, one can set up a hierarchy of non-linear evolution equations. This is attained by using recurrence relations which enable construction of the hierarchy through use
  of certain polynomial functions, generating the Lax pairs and thereby the different equations of the hierarchy. One of the ongoing challenges in the study of integrable systems is to construct such systems associated with non-linear evolution equations of physical significance.

\bigskip

\paragraph*{} Two relatively recent developments in the field of integrable systems seem to be having an important impact on the growth and orientation of the field. The first one was initiation of the concept of non-holonomic deformation of integrable models in 2008 while the second event was introduction of non-local integrable systems in 2013. Considering the latter one first, we note that the Nonlinear Schr\"odinger (NLS) equation, in one space and one time (1+1) variables is an extremely well known classical nonlinear integrable equation, appearing in diverse physical systems \cite{Malo, Zhid} encompassing nonlinear optics, plasma physics, fluid mechanics and also in mathematical fields like differential geometry \cite{RS}. A new integrable reduction of this well known equation was obtained \cite{AM1}, called the Non-local Nonlinear Schrödinger Equation. This equation is PT symmetric, \cite{Bender1, Bender2, Bender3} i.e. it is symmetric under the parity -time transfor
 m $x\rightarrow -x$, $t\rightarrow -t$, $q\rightarrow q^{*}$. It has been introduced as a mathematical model to discuss wave propogation in PT symmetric nonlinear media \cite{Guo, Regens, Ruter}. A multi-dimensional analogue of the non-local nonlinear Schrödinger equation was also discovered \cite{Fok}. A variety of integrable non-local nonlinear equations were obtained through symmetry reductions of the AKNS scattering problem, the non-locality appearing in both space and time or time alone \cite{AM2}. Examples include, apart from nonlocal NLS, the modified KdV, sine-Gordon, $(1+1)$ and $(2+1)$ dimensional three-wave interaction, derivative NLS, Davey-Stewartson etc. Several related work were undertaken subsequently\cite{AM3, AM4, AM5, Val,Gurses,GS}which dealt with different aspects of nonlocal integrable systems.\\

\noindent
The second development, viz. the non-holonomic deformation of integrable systems is one in which the system is perturbed in such a manner that under suitable differential constraints on the perturbing function, the system retains its integrability. It was shown by Karasu-Kalkani {\it et al} \cite{KK} that the integrable 6th order KdV equation represented a nonholonomic deformation(NHD) of the celebrated KdV equation. The terminology ``nonholonomic deformation" was used by Kupershmidt \cite{Kuppershmidt}. In \cite{AK1} a matrix Lax pair, the N-soliton solution using inverse scattering transform and a two-fold integrable hierarchy were obtained for the non-holonomic deformation of the KdV equation. In \cite{AK2} the work was extended to include the non-holonomic deformation of both KdV and mKdV equations along with their symmetries, hierarchies and integrability. The non-holonomic deformation of derivative NLS and Lenells-Fokas equations was discussed in \cite{AK3}, while such
  deformation of generalized KdV type equations was taken up in \cite{Guha1} where emphasis was put on the geometrical aspect of the problem. Kupershmidt's infinite-dimensional construction was extended in  \cite{Guha2} to obtain non-holonomic deformation of a wide class of coupled KdV systems, all of which are generated from the Euler-Poincare-Suslov flows. A comparative study encompassing two different types of deformations, viz. non-holonomic and quasi-integrable, of equations in the NLS and DNLS hierarchy was undertaken in \cite{AGM}.   \\

\subsection{The purpose, result and structure of the paper}
Consider a coupled system of evolution equations given by
\beq
iq_t = q_{xx} - 2q^2r, \qquad -ir_t = r_{xx} - 2r^2q,
\eeq
where $q(x, t)$ and $r(x, t)$ are potentials of the well-known AKNS $2 \times 2$ linear scattering
problem. We obtain the celebrated NLS equation when $ r = \sigma q^{\ast}$, $\sigma = \mp{1}$. As discussed above,
Ablowitz and Musslimani \cite{AM1} obtained a new interesting reduction of the
AKNS scattering problem, given by $r(x, t) = \sigma q^{*} (-x, t)$. This leads to
the integrable nonlocal NLS equation having profound applications in
PT symmetric quantum physics and optics, giving rise to the name PT symmetric NLS equation (or PTNLS in short).  Ablowitz and Musslimani \cite{AM1} developed the inverse scattering transform for decaying data and obtained a one breathing soliton solution.

More recently, Ablowitz and Musslimani \cite{AM2} found two
more reductions of the AKNS scattering problem leading to interesting nonlocal NLS
type equations. These are given by $r(x,t) = \sigma q(-x, -t)$ and $r(x, t) = \sigma q(x, -t)$,
and the equations which are obtained from these two reductions are called
reverse space-time NLS (RSTNLS) and reverse time NLS (RTNLS) equations respectively.
Ablowitz and Musslimani have found many
other nonlocal integrable equations such as nonlocal modified Korteweg-de Vries equation, nonlocal Davey-Stewartson equation, nonlocal sine-Gordon equation, and nonlocal (2 + 1)-dimensional three-wave interaction equations. The work carried out by Ablowitz and Musslimani has led to a flurry of activities on nonlocal reductions of systems of integrable equations.

\smallskip

The purpose of the present work is to tie together the two relatively recent developments in the field of integrable systems by examining the effect of non-holonomic deformation on non-local integrable systems. The non-local systems are obtained by imposing suitable symmetry reductions on one of the dynamical variables occurring in the Lax pair. On the other hand, the deformation could be achieved either starting from the bi-Hamiltonian structure of the system of equations following the method adopted in \cite{Kuppershmidt} or by inserting perturbing functions in the temporal component of the Lax pair. The study is undertaken in respect of the NLS and the DNLS equations by first obtaining the corresponding non-local systems and then deforming the systems so obtained. The latter is implemented using both approaches mentioned above. The NLS equation is chosen as a test platform in view of its generic nature and wide range of applications.\\

The {\bf results} obtained indicate the following:\\
(i) non-locality gets introduced not only in the dynamical variables but also in the perturbing variables;\\
(ii)the perturbing variables enter as "source" terms and make the deformed equations inhomogenous compared to the original integrable systems;\\
(iii)the bi-hamiltonian approach shows that the constraints on the deforming functions are integro-differential in nature;\\
(iv)an exact equivalence is easy to establish between the bi-hamiltonian and Lax pair approaches in case of the Nonlinear Schr\"odinger (NLS) equation. The results become more complex in case of the Kaup-Newell system of equations.      

\smallskip

\paragraph*{}The rest of the paper is {\bf organized} as follows: Section 2 discusses the Non-local Nonlinear Schrödinger (NNS)equation and its bi-Hamiltonian structure. Section 3 demonstrates the Non-Holonomic Deformation of the NNS system by using the Lax pair approach as well as the Kupershmidt prescription and establishes the equivalence of the two approaches. Section 4 considers the derivative NLS using the Kaup-Newell equation and derives the non-local counterpart. Section 5 is devoted to the study of non-holonomic deformtion of the non-local Kaup-Newell system. Section 6 lists possible outcomes and indicates how the study may be carried forward.\\

\section{Non-local Nonlinear Schr\"odinger equation and its bi-Hamiltonian structure}

We start with the Lax pair of the coupled Nonlinear Schr\"odinger equations given by
\bqa
&& U = -i \lambda \sigma_3 +q (x,t) \sigma_+ + r (x,t) \sigma_-,  \nonumber\\
&& V = \left(2i\lambda^2+iq (x,t)r(x,t)\right)\sigma_3 -2 \lambda(q (x,t)\sigma_+ \nonumber\\
&& + r (x,t)\sigma_-) + i\left(-q_x (x,t)\sigma_+ +r_x (x,t)\sigma_-\right). \label{N06}
\eqa

The coupled NLS equations, in terms of both $q$ and $r$ are given by
\beq
q_t = -iq_{xx}+ 2iq^2r,\qquad r_t = ir_{xx}-2ir^2q;
\eeq
and can be obtained as consistency equations, by imposing the usual zero-curvature condition:
\beq
U_t-V_x+[U,V]=0. \label{N02}
\eeq
The bi-Hamiltonian structures of the pair of NLS equations are given by:

\bqa
			&& B^1=  \left(\begin{array}{cc}
0 & -i \\
                                           i  & 0  \\
                                    \end{array}\right),\nonumber\\
&& B^2 = \left(\begin{array}{cc}
q\partial_x^{-1}q & \frac{1}{2}\partial_x - q\partial_x^{-1}r \\
                                            \frac{1}{2}\partial_x - r\partial_x^{-1}q &  r\partial_x^{-1}r \\
                                    \end{array}\right),
\eqa
and the corresponding conserved densities are:
\beq
H^1 = -\int(q_xr_x+ q^2r^2)dx \quad{\rm and}\quad H^2 = i\int(r_xq - q_xr)dx.
\eeq

\paragraph*{} Under the standard AKNS symmetry, we put
\beq
r(x,t) = \sigma q^{*}(x,t)
\eeq

On putting $\sigma = 1$, equation (3) reduces to\\
\beq
q_t(x,t) = -iq_{xx}(x,t)+ 2iq^2(x,t)q^{*}(x,t),\qquad q^{*}(x,t)_t = iq^{*}_{xx}(x,t)-2iq^{*2}(x,t)q(x,t)
\eeq

Under this symmetry, the Hamiltonian structures are

\bqa
			&& B^1=  \left(\begin{array}{cc}
0 & -i \\
                                           i  & 0  \\
                                    \end{array}\right),\nonumber\\
&& B^2 = \left(\begin{array}{cc}
q\partial_x^{-1}q & \frac{1}{2}\partial_x - q\partial_x^{-1}q^{*} \\
                                            \frac{1}{2}\partial_x - q^{*}\partial_x^{-1}q &  q^{*}\partial_x^{-1}q^{*} \\
                                    \end{array}\right),
\eqa

while the conserved densities become
\beq
H^1 = -\int(q_xq^{*}_x+ q^2q^{*2})dx \quad{\rm and}\quad H^2 = i\int(q^{*}_xq - q_xq^{*})dx.
\eeq
\paragraph*{} Under the parity-time (PT) preserving symmetry which generates the non-local non-linear Schr\"odinger equation, we take
\beq
r(x,t) = q^{*}(-x,t)
\eeq
whereby (3) gives the pair of non-local non-linear Schr\"odinger equations (NNS), viz.

\beq
q_t(x,t) = -iq_{xx}(x,t)+ 2iq^2(x,t)q^{*}(-x,t),\qquad q^{*}(-x,t)_t = iq^{*}_{xx}(-x,t)-2iq^{*2}(-x,t)q(x,t)
\eeq

The foregoing equations are non-local since the evolution of the dynamical variable at the transverse coordinate x always requires information from the opposite point $(-x)$. \\

To check $PT$ symmetry, the first of the equations in $(12)$ is rewritten as
\beq
iq_t(x,t) = q_{xx}(x,t)+ V (q,x,t)q(x,t)
\eeq

where
\beq
V(q,x,t) = -2q(x,t)+ q^{*}((-x,t)
\eeq
is the self-induced potential. \\
It is easy to show that
\beq
V(q,x,t) = V^{*}(q,-x,t)
\eeq
which establishes the $PT$ symmetry condition. \\

The bi-Hamiltonian structure of the pair of $NNS$ equations are given by

\bqa
			&& B^1_{NNS}=  \left(\begin{array}{cc}
0 & -i \\
                                           i  & 0  \\
                                    \end{array}\right),\nonumber\\
&& B^2_{NNS} = \left(\begin{array}{cc}
q(x,t)\partial_x^{-1}q(x,t) & \frac{1}{2}\partial_x - q(x,t)\partial_x^{-1}q^{*}(-x,t) \\
                                            \frac{1}{2}\partial_x - q^{*}(-x,t)\partial_x^{-1}q(x,t) &  q^{*}(-x,t)\partial_x^{-1}q^{*}(-x,t) \\
                                    \end{array}\right),
\eqa
Here the arguments of $q$ and $q^{*}$ are written out explicitly to emphasize the non-local character of the equations. \\

The corresponding conserved densities are
\beq
H^1_{NNS} = -\int(q_xq^{*}_x+ q^2q^{*2})dx,\qquad H^2_{NNS} = i\int(q^{*}_x q - q_x q^{*})dx
\eeq

As before $q = q(x,t)$  and $q^{*} = q^{*}(-x,t)$ in the immediately preceding equation.

\section{Non-holonomic deformation of Non-local Nonlinear Schr\"odinger equation}
\subsection{The Bi-Hamiltonian approach}
In this formalism, one uses $(i)$ the bi-Hamiltonian structure of the parent equation and $(ii)$ deforming variables, to obtain the deformed equations as well as the constraint conditions on the deforming functions. In the present case, the pair of deformed non-local non-linear Schr\"odinger equations may be expressed as

\bqa
\left(
\begin{array}{cc}
q (x,t)\\
q^{*}(-x,t)
\end{array}
\right)_t&=&B^1_{NNS}\left(
\begin{array}{cc}
\frac{\delta}{\delta q}\\
\frac{\delta}{\delta q^{*}}
\end{array}
\right)H^1_{NNS}-B^1_{NNS}\left(
\begin{array}{cc}
g(x,t)\\
g^{*}(-x,t)
\end{array}
\right)\nonumber\\
&=&B^2_{NNS}\left(
\begin{array}{cc}
\frac{\delta}{\delta q}\\
\frac{\delta}{\delta q^{*}}
\end{array}
\right)H^2_{NNS}-B^1_{NNS}\left(
\begin{array}{cc}
g(x,t)\\
g^{*}(-x,t)
\end{array}
\right),
\eqa

where $g(x,t)$ and $g^{*}(-x,t)$ are the perturbing functions, $*$ denotes the complex conjugate, and the argument of the complex conjugate is taken suitably to fit into the $NNS$ scheme.\\

Following the method adopted in \cite{Kuppershmidt} $(18)$ yields the following pair of deformed non-local non-linear Schrödinger equations\\
\beq
q_{t}(x,t) = -iq_{xx}(x,t) + 2iq^{*}(-x,t)q^{2}(x,t) + ig^{*}(-x,t)
\eeq
\beq
q^{*}_{t}(-x,t) = iq^{*}_{xx}(-x,t) - 2iq^{*2}(-x,t)q(x,t) - ig(x,t)
\eeq

The constraints on the perturbing variables $g(x,t)$ and $g^{*}(-x,t)$ are obtained by setting \\
\beq
B^2_{NNS}\left(
\begin{array}{cc}
g(x,t)\\
g^{*}(-x,t)
\end{array}
\right)=0,
\eeq
that leads to the conditions:\\
\beq
g^{*}_{x}(-x,t) + 2 q(x,t)\partial^{-1} [q(x,t)g(x,t) - q^{*}(-x,t)g^{*}(-x,t)] = 0
\eeq

\beq
g_{x}(x,t) + 2 q^{*}(-x,t)\partial^{-1} [q^{*}(-x,t)g^{*}(-x,t) - q(x,t)g(x,t)] = 0
\eeq

Thus the constraints, as they stand, are integro-differential in nature. \\

Multiplying $(22)$ by $q^{*}(-x,t)$ and $(23)$ by $q(x,t)$ and adding, one obtains\\
\beq
q(x,t)g_x(x,t) + q^{*}(-x,t)g^{*}_x(-x,t) = 0
\eeq
\subsection{Lax pair approach}
To construct the NHD in this method, one starts with a Lax pair, keeping the space part $U(\lambda)$ unchanged but modifying the temporal component $V(\lambda)$. This implies that the scattering problem remains unchanged, but the time evolution of the spectral data becomes different in the perturbed models. To retain integrability the non-holonomic constraints have to be affine in velocities prohibiting explicit velocity dependence of the deformed dynamical system. This insists on deformation of the temporal component of the Lax pair only as in absence of its time derivative in the flatness condition the dynamical equation can remain velocity independent \cite{NHD1}. It is due to such construction that the system can retain its integrability inspite of being subject to perturbation. \\
\paragraph{} Under the $PT$ symmetric transformation, the Lax pair takes the form
\beq
U = -i\lambda\sigma_3 + q(x,t)\sigma_+ q^{*}(-x,t)\sigma_-
\eeq
\beq
V_{original}= [2i\lambda^{2} + iq(x,t)q^{*}(-x,t)]\sigma_3 - [2\lambda q(x,t)+ iq_x(x,t)]\sigma_+ + [-2\lambda q^{*}(-x,t) + iq^{*}_x (-x,t)]\sigma_-
\eeq
where $V_{original}$ is the undeformed temporal component of the Lax pair and the arguments of the variables have been explicitly shown to emphasize the non-locality present. \\
The only scale present in the system is the spectral parameter $\lambda$, defining the corresponding solution space. In order to obtain
a deformation of the temporal part, that maintains integrability through the flatness condition of the type in equation $(4)$, it is intuitively obvious that the deformation part will be a function of $\lambda$. We propose the following additive
deformation term to the temporal Lax component of the non-local NLS equation\\

\beq
V_{deformation} = \frac{1}{2}\lambda^{-1} [a(x,t)\sigma_3 + b(x,t)\sigma_+ + c(x,t)\sigma_-]
\eeq
The adopted deformation of equation $(27)$ contains only ${\cal O}\left(\lambda^{-1}\right)$ terms. The presence of higher
order terms in $\lambda$ (zero or positive powers) only lead to additional perturbed dynamical systems at each order, and
vanish when the terms are substituted order-by-order. This is because the non-local NLS equations arise from contributions of
${\cal O}\left(\lambda^{0}\right)$ {\it in} the flatness condition, and the presence of any higher order contribution is
decoupled from the dynamics governed by the same. Therefore, the highest order deformations end-up yielding trivial identities,
that eventually eliminates all the contribution with positive powers of $\lambda$ in the non-local NLS equation. One can verify this
by adding a term with positive powers of $\lambda$ to $V_{\rm deformed}$. Therefore, the expression in equation $(27)$ is
general as far as zero or positive powers of $\lambda$ are concerned.

Taking $\tilde{V} = V_{original} + V_{deformation}$ and imposing the zero curvature or flatness condition, the non-holonomic deformed equations are obtained as the ${\cal O}\left(\lambda^{0}\right)$ terms in the above condition
\beq
q_t(x,t)+ iq_{xx}(x,t) - 2iq^{2}(x,t)q^{*}(-x,t) - ib = 0
\eeq
\beq
q^{*}_t(-x,t)- iq^{*}_{xx}(-x,t) + 2iq^{*2}(-x,t)q(x,t) + ic = 0
\eeq
Putting $b(x,t) = g^{*}(-x,t)$ and $c(x,t) = g(x,t)$ in equations $(28)$ and $(29)$ respectively, we are led to the non-holonomic deformed coupled non-local NLS equations\\
\beq
q_t(x,t)= -iq_{xx}(x,t) + 2iq^{2}(x,t)q^{*}(-x,t) + ig^{*}(-x,t)
\eeq
\beq
q^{*}_t(-x,t)= iq^{*}_{xx}(-x,t) - 2iq^{*2}(-x,t)q(x,t) - ig(x,t)
\eeq
Such deformations are trivially expected and they effectively make the original non-local NLS equations inhomogeneous by introducing
{\it source terms} $g$ and $g^{*}$ to the dynamics. Such equations are already known to form integrable systems, and thus, satisfy
our primary goal.\\
The differential constraints con the perturbing variables are obtained from the ${\cal O}\left(\lambda^{-1}\right)$ sector in the flatness condition, on equating the coefficients of the generators $\sigma_3$, $\sigma_+$ and $\sigma_-$ :\\
\beq
a_x = q(x,t)g(x,t) - q^{*}(-x,t)g^{*}(-x,t)
\eeq
\beq
g^{*}_x(-x,t) + 2q(x,t)a(x,t) =0
\eeq
\beq
g_x(x,t) - 2q^{*}(-x,t)a(x,t) = 0
\eeq

Eliminating $a(x,t)$ from the above equations, we obtain a single differential constraint as\\
\beq
q(x,t)g_{xx}(x,t) + q^{*}_x(-x,t) g^{*}_x (-x,t) + 2q(x,t)q^{*}(-x,t)[q^{*}(-x,t)g^{*}(-x,t) - q(x,t)g(x,t)] = 0
\eeq

Using $(30)$ and $(31)$ to eliminate the remaining perturbing variables $g(x,t)$ and $g^{*}(-x,t)$ from equation $(35)$, we derive a single higher order  non-holonomic deformed non-local Non-linear Schr\"odinger equation given as:
\bqa
&&q(x,t) \Big[\partial^{2} - 2q(x,t)q^{*}(-x,t)\Big]\Big[iq^{*}_t(-x,t)+ q^{*}_{xx}(-x,t)- 2q(x,t)q^{*2}(-x,t)\Big]  \nonumber\\
 && + q^{*}_{x}(-x,t)\Big[-iq_t(x,t) + q_{xx}(x,t)- 2q^{2}(x,t)q^{*}(-x,t)\Big]_{x} \nonumber\\
&& + 2q(x,t)q^{*2}(-x,t)\Big[-iq_t(x,t) + q_{xx}(x,t) - 2q^{2}(x,t)q^{*2}(-x,t)\Big]. \label{N06}
\eqa

Such an equation is subjected to the dynamics of equations $(30)$ and $(31)$, and therefore do not yield any new dynamics, and eventually
reflects the constraint itself in a different form. This is in accord with the previous argument that no term, with power of
$\lambda$ other than that responsible for yielding equations $(30)$ and $(31)$, can yield dynamics to the non-local NLS system, as it will violate the overall integrability of the system itself.\\
The constraint of equation $(35)$ is {\it non-holonomic} in nature, as it contains differentials of corresponding variables,
and characterizes the corresponding deformation. It is crucial that such a constraint solely arises from the terms with negative power of the spectral parameter. Further, explicit forms of the local functions $a$, $g_1$ and $g_2$ are not necessary to establish the integrability, and they represent a class that satisfies the constraint in equation $(35)$. In other words, the constraints arise form ${\cal O}\left(\lambda^{-1}\right)$ contributions, and {\it additionally}
restrict the allowed values of $q(x,t)$ and $q^{*} (-x, -t)$ of the deformed dynamics of ${\cal O}\left(\lambda^0\right)$.

From  $(33)$ and $(34)$ we obtain
\beq
q^{*}(-x,t)g^{*}_{x}(-x,t) + q(x,t)g_x(x,t) = 0
\eeq
which agrees with $(24)$, thereby establishing the equivalence of the Bi-Hamiltonian and Lax pair approaches.

\section{Derivative Nonlinear Schr\"odinger equations and their bi-Hamiltonian structure}
\subsection{The coupled Kaup Newell system}

We consider the Lax pair of the Kaup-Newell (KN)system which belongs to the Derivative Nonlinear Schr\"odinger family of integrable systems. This is given by\\
\beq
U = -i\lambda^{2}\sigma_3 + \lambda q\sigma_+ + \lambda r\sigma_-
\eeq
\beq
V = -i(2\lambda^{4} +\lambda^{2}qr)\sigma_3 + \Big[2\lambda^{3}q + \lambda (iq_x + q^{2}r)\Big]\sigma_+ + \Big[2\lambda^{3}r + \lambda (-ir_x + qr^{2})\Big]\sigma_-
\eeq
The zero curvature condition leads to the following coupled Kaup-Newell system of equations\\
\beq
q_t(x,t)= iq_{xx}(x,t) + (q^{2}(x,t)r(x,t))_x
\eeq
\beq
r_t(x,t)= -ir_{xx}(x,t) + (q(x,t)r^{2}(x,t))_x
\eeq

The bi-Hamiltonian structure of the Kaup-Newell system is given by

\bqa
			&& B^1_{KN}=  \left(\begin{array}{cc}
0 & \partial \\
                                           \partial  & 0  \\
                                    \end{array}\right),\nonumber\\
&& B^2_{KN} = B^1_{KN} J_{KN} B^1_{KN}
\eqa

where
\bqa
&& J =  \left(\begin{array}{cc}
                                    r\partial^{-1}r & \-i + r\partial^{-1}q \\
                                            \ i + q\partial^{-1}r &  q\partial^{-1}q \\

                                    \end{array}\right),
\eqa
The conserved densities are		
\beq
H^1_{KN} = \frac{1}{2}\int(q^{2}r^{2}+ i(q_{x}r - qr_{x}))dx \quad{\rm and}\quad H^2 = \int(qr)dx.
\eeq

\subsection{Symmetry reduction and non-local coupled Kaup-Newell system}
We impose the symmetry reduction\\
\beq
r(x,t) = q(-x, -t)
\eeq
Under this reduction the system of equations $(40)$ and $(41)$ are compatible and we obtain the reverse space-time non-local coupled "classical" $DNLS$ equations or the non-local coupled Kaup-Newell system given below:\\
\beq
q_{t}(x,t) = iq_{xx}(x,t) + (q^{2}(x,t)q(-x, -t))_x
\eeq
\beq
q_{t}(-x,-t) = -iq_{xx}(-x,-t) + (q(x,t)q^{2}(-x, -t))_x
\eeq
The non-local character of the equations arises from the fact that evolution of the field at a given spatial and temporal coordinate $x$ and $t$ is governed not only by the values of the field at these coordinates, but also by the information coming from the corresponding reversed coordinates viz. $(-x)$ and $(-t)$. \\
The bi-Hamiltonian structure of the above equations are given by\\
\bqa
			&& B^1_{NKN}=  \left(\begin{array}{cc}
0 & \partial \\
                                           \partial  & 0  \\
                                    \end{array}\right),\nonumber\\
&& B^2_{NKN} = B^1_{NKN} J_{NKN} B^1_{NKN}
\eqa

where
\bqa
&& J_{NKN} =  \left(\begin{array}{cc}
                                    q(-x, -t)\partial^{-1}q(-x, -t) & \-i + q(-x, -t)\partial^{-1}q(x,t) \\
                                            \ i + q(x,t)\partial^{-1}q(-x, -t) &  q(x,t)\partial^{-1}q(x,t) \\

                                    \end{array}\right),
\eqa
Here $NKN$ denotes the non-local Kaup-Newell system. \\
The conserved densities are given by \\
\bqa
&& H^1_{NKN} = \frac{1}{2}\int(q^{2}(x,t)q^{2}(-x,-t)+ i(q_{x}(x,t)q(-x,-t) - q(x,t)q_{x}(-x,-t)))dx  \nonumber\\
&& H^2_{NKN} = \int(q(x,t)q(-x,-t))dx \label{N06}
\eqa

\section{Non-holonomic deformation of non-local coupled Kaup Newell system}
The methods and arguments adopted while deriving the non-holonomic deformation of the non-local coupled Kaup Newell system are similar to those followed in case of the non-local coupled Nonlinear Schr\"odinger equations.
\subsection{The Bi-Hamiltonian approach}
In this formalism, the non-local coupled Kaup-Newell systems, under non-holonomic deformation, are expressed in terms of the bi-Hamiltonian structure and the conserved densities as \\
\bqa
\left(
\begin{array}{cc}
q (x,t)\\
q(-x,-t)
\end{array}
\right)_t&=&B^1_{NKN}\left(
\begin{array}{cc}
\frac{\delta}{\delta q(x,t)}\\
\frac{\delta}{\delta q(-x, -t)}
\end{array}
\right)H^1_{NKN}-B^1_{NKN}\left(
\begin{array}{cc}
g^{1}(x,t)\\
g^{2}(-x,-t)
\end{array}
\right)\nonumber\\
&=&B^2_{NKN}\left(
\begin{array}{cc}
\frac{\delta}{\delta q(x,t)}\\
\frac{\delta}{\delta q(-x, -t)}
\end{array}
\right)H^2_{NKN}-B^1_{NKN}\left(
\begin{array}{cc}
g^{1}(x,t)\\
g^{2}(-x,-t)
\end{array}
\right),
\eqa
Here $g^{1}(x,t)$ and $g^{2}(-x,-t)$ are the perturbing functions; the argument of the second function is suitably modified to fit the reverse space-time non-local character of the parent equations. \\
From $(51)$, the pair of deformed non -local coupled $KN$ equations are\\
\beq
q_{t}(x,t) = iq_{xx}(x,t) + (q^{2}(x,t)q(-x,-t))_x - g^{2}_x(-x, -t)
\eeq
\beq
q_{t}(-x,-t) = -iq_{xx}(-x,-t) + (q^{2}(-x,-t)q(x,t))_x - g^{1}_x(x, t)
\eeq
The constraints on the perturbing variables $ g^{1}(x, t)$ and $g^{2}(-x, -t)$ are obtained by setting\\
\beq
B^2_{NKN}\left(
\begin{array}{cc}
g^{1}(x,t)\\
g^{2}(-x,-t)
\end{array}
\right)=0,
\eeq
and yield the following equations\\
\beq
ig^{2}_{xx}(-x, -t) + (q(x,t)B)_{x} = 0
\eeq
\beq
-ig^{1}_{xx}(x,t) + (q(-x,-t)B)_{x} = 0
\eeq
where $B = \partial^{-1}A$, and
\beq
A = q(x,t)g^{1}_{x}(x,t) + q(-x, -t)g^{2}_{x}(-x, -t) = 0
\eeq
\subsection{Lax pair approach}
To derive the $non-holonomic$ deformation in case of the non-local Kaup Newell system, one starts with the following Lax pair\\
\beq
U = -i\lambda^{2}\sigma_{3} + \lambda q(x,t)\sigma_{+} + \lambda q(-x, -t)\sigma_{-}
\eeq
\bqa
&& V_{original}= (-2i\lambda^{4}- i\lambda^{2} q(x,t)q(-x, -t))\sigma_{3}  \nonumber\\
 && + (2\lambda^{3}q(x,t) + \lambda (iq_x(x,t) + q^{2}(x,t)q(-x, -t)))\sigma_{+} \nonumber\\
&& + (2\lambda^{3}q(-x, -t) + \lambda (-iq_x(-x, -t)+ q(x,t)q^{2}(-x, -t)))\sigma_{-} \label{N06}
\eqa
where $V_{original}$ denotes the undeformed temporal component of the Lax pair generating the non-local coupled Kaup-Newell system of equations. \\

\paragraph*{} We introduce deformation in the time component above by defining\\
\beq
V_{deformed} = \alpha (G^{0} + \lambda^{-1} G^{1} + \lambda^{-2} G^{2})
\eeq
where \\
\beq   \begin{array}{lc}
G^{(0)} = w \sigma_3 + m_1 \sigma_+ + m_2 \sigma_-
\end{array}   \eeq
\beq   \begin{array}{lc}
G^{(1)} = a \sigma_3 + g_1 \sigma_+ + g_2 \sigma_-
\end{array}   \eeq
\beq   \begin{array}{lc}
G^{(2)} = b \sigma_3 + f_1 \sigma_+ + f_2 \sigma_-
\end{array}   \eeq
\paragraph*{} Taking $\tilde{V} = V_{original} + V_{deformed}$ and using the zero curvature condition on U defined by $(58)$ and $\tilde{V}$, we arrive at the following results for the variables contained in the functions $G^{0}$, $G^{1}$ and $G^{2}$:\\
$m_{1} = 0$, $m_{2} = 0$, $a = 0$, $f_{1} = 0$, $f_{2} = 0$ and $b = b(t)$ where the last entry indicates that b is a function of t only.  \\
The pair of non-holonomic deformed coupled non-local Kaup-Newell equations are given below\\
\beq
q_t(x,t) - iq_{xx}(x,t) - (q^{2}(x,t) q(-x, -t))_x - 2i\alpha g_1 - 2\alpha w q(x,t) = 0
\eeq
\beq
q_t(-x,-t) + iq_{xx}(-x,-t) - (q^{2}(-x,-t) q(x, t))_x + 2i\alpha g_2 - 2\alpha w q(-x,-t) = 0
\eeq
The differential constraints on the non-zero variables $w$, $g_{1}$ and $g_{2}$ are given by\\
\beq
w_x = q(x,t)g_2 - q(-x, -t)g_1
\eeq
\beq
g_{1x} + 2b(t) q(x,t)=0
\eeq
\beq
g_{2x} - 2b(t) q(-x,-t)=0
\eeq
We now try to obtain new non-linear integrable systems by resolving the constraint relations and expressing the perturbing variables in terms of the basic field variables. To this end, let us put \\
\bqa
&& q(x,t)= u_x(x,t)  \nonumber\\
&& q(-x,-t) = v_x (-x, -t) \label{N06}
\eqa
Using $(69)$ in $(66)$,$(67)$ and $(68)$ we are led to the following relations:\\
\bqa
&& g_1 = -2b(t) u(x,t)  \nonumber\\
&& g_2 = 2b(t)v(-x, -t) \nonumber\\
&& w = 2b(t)u(x,t)v(-x, -t) + K(t) \label{N06}
\eqa
Eliminating $g_1$, $g_2$ and $w$ from $(64)$ and $(65)$ by using $(70)$, we can rewrite the non-holonomic deformed coupled non-local (space and time reversed) Kaup-Newell system as \\
\bqa
&& u_{xt}(x,t) - iu_{xxx}(x,t) - (u_{x}^{2}(x,t)v_x(-x, -t))_x + 4i\alpha b(t)u(x,t)  \nonumber\\
&& - 2\alpha u_x(x,t)(2b(t)u(x,t)v(-x, -t) +K(t))= 0 \label{N06}
\eqa

\bqa
&& v_{xt}(-x,-t) + iv_{xxx}(-x,-t) - (u_{x}(x,t)v_x^{2}(-x, -t))_x + 4i\alpha b(t)v(-x,-t)  \nonumber\\
&& + 2\alpha v_x(-x,-t)(2b(t)u(x,t)v(-x, -t) +K(t))= 0 \label{N06}
\eqa
\paragraph*{} Equations $(71)$ and $(72)$ are coupled non-local evolution equations that are non-autonomous in character with arbitrary time dependent coefficients $b(t)$ and $K(t)$. The constraints have been resolved and thus removed from the system. These equations may be regarded as generalized non-local versions of the Lennels-Fokas equations with the inclusion of a non-linear derivative term as well as higher order dispersion term.

\section{Discussion and conclusion} The study of non-local integrable systems is a relatively new area in the domain of integrable systems.
The results in this paper are significant in two different
ways. Firstly, from a mathematical point of view, we presented
integrable nonlocal equations which have some physical applications in PT-symmetric quantum physics. The bi-hamiltonian structure of these nonlocal equations have been clearly furnished as part of the analysis. Secondly, this is probably the first work in which such systems have been studied under the lens of non-holonomic deformation, which is a perturbing method that retains integrability of the system. The deformation is carried out using the bi-Hamiltonian formalism due to Kupershmidt as well as the Lax pair approach. The systems chosen were those belonging to the Nonlinear Schr\"odinger family, since the Nonlinear Schr\"odinger equation is one of the most celebrated equations having a number of physical applications.

Our next programme will be to study the soliton nd multi-solitons solutions of these
new set of nonlocal NLS equations. It will also be interesting to study other aspects of non-local systems ofother members of the Nonlinear Schr\"odinger equation family,
such as their behaviour under Quasi-integrable deformation (QID) \cite{f1, New1} obtaining them under reduction of the Self-dual Yang Mills (SDYM) equations \cite{ACT, ACH} and their connection with other integrable systems \cite{AM6}. These topics will form the basis of our future investigations in this area \cite{MGA New}. \\

\section*{Acknowledgements}
We would like to express our sincere appreciation to Professor Sarbarish Chakravarty for his encouragement and  enlightening discussions.We would also like to thank Dr. Kumar Abhinav for his interest.

\section*{An observation}
A large number of papers have been published on non-local integrable systems in the past few years. We have cited only those papers which have a direct bearing on and relevance to the present work.

\end{document}